\documentclass[pra,aps,preprint,superscriptaddress,showpacs,nofootinbib]{revtex4-1}

\usepackage{layout}
\usepackage{color}
\usepackage[utf8]{inputenc}
\usepackage{amsmath}
\usepackage{tikz}
\usepackage{xcolor}
\usetikzlibrary{arrows}

\usepackage{subfigure, epsfig}
\usetikzlibrary[topaths]

\usepackage{changes}
\usepackage{amsfonts}
\usepackage{amsmath}
\usepackage{amsthm}
\usepackage{amscd}
\usepackage{amssymb}
\usepackage{amsxtra}
\usepackage{bm}
\usepackage{bbm}
\usepackage{exscale}
\usepackage{graphics,color}
\usepackage{latexsym}
\usepackage{enumerate}
\usepackage{dcolumn}
\usepackage{graphics}
\usepackage{epstopdf}
\usepackage{graphicx}
\usepackage{hyperref}
\hypersetup{
    colorlinks,%
    citecolor=black,%
    filecolor=black,%
    linkcolor=black,%
    urlcolor=black
}

\newcommand{\ketbra}[2]{\ensuremath{|#1\rangle\langle #2|}}

\newcommand{\beq}{\begin{equation}}
\newcommand{\eeq}{\end{equation}}

\newcommand{\bdf}{\begin{defn}}
\newcommand{\edf}{\end{defn}}

\newtheorem{defn}{Definition}

\newcount\mycount
\begin{document}

\title{\textbf{Experimental unconditionally secure covert communication in dense wavelength-division multiplexing networks}}

\author{Yang Liu}
\affiliation{Shanghai Branch, National Laboratory for Physical Sciences at Microscale and Department of Modern Physics, University of Science
and Technology of China, Shanghai, 201315, P.~R.~China}
\affiliation{CAS Center for Excellence and Synergetic Innovation Center in Quantum Information and Quantum Physics, University of Science and
Technology of China, Shanghai 201315, P.~R.~China}
\author{Juan Miguel Arrazola}
\affiliation{Centre for Quantum Technologies, National University of Singapore, 3 Science Drive 2, Singapore 117543}
\author{Wen-Zhao Liu}
\affiliation{Shanghai Branch, National Laboratory for Physical Sciences at Microscale and Department of Modern Physics, University of Science
and Technology of China, Shanghai, 201315, P.~R.~China}
\affiliation{CAS Center for Excellence and Synergetic Innovation Center in Quantum Information and Quantum Physics, University of Science and
Technology of China, Shanghai 201315, P.~R.~China}
\author{Weijun Zhang}
\affiliation{State Key Laboratory of Functional Materials for Informatics, Shanghai Institute of Microsystem and Information Technology, Chinese Academy of Sciences, Shanghai 200050, P.~R.~China}
\author{Ignatius William Primaatmaja}
\affiliation{Department of Physics, National University of Singapore, 2 Science Drive 3, Singapore 117542.}
\author{Hao Li}
\author{Lixing You}
\author{Zhen Wang}
\affiliation{State Key Laboratory of Functional Materials for Informatics, Shanghai Institute of Microsystem and Information Technology, Chinese Academy of Sciences, Shanghai 200050, P.~R.~China}
\author{Valerio Scarani}
\affiliation{Centre for Quantum Technologies, National University of Singapore, 3 Science Drive 2, Singapore 117543}
\affiliation{Department of Physics, National University of Singapore, 2 Science Drive 3, Singapore 117542.}
\author{Qiang Zhang}
\affiliation{Shanghai Branch, National Laboratory for Physical Sciences at Microscale and Department of Modern Physics, University of Science
and Technology of China, Shanghai, 201315, P.~R.~China}
\affiliation{CAS Center for Excellence and Synergetic Innovation Center in Quantum Information and Quantum Physics, University of Science and
Technology of China, Shanghai 201315, P.~R.~China}
\author{Jian-Wei Pan}
\affiliation{Shanghai Branch, National Laboratory for Physical Sciences at Microscale and Department of Modern Physics, University of Science
and Technology of China, Shanghai, 201315, P.~R.~China}
\affiliation{CAS Center for Excellence and Synergetic Innovation Center in Quantum Information and Quantum Physics, University of Science and
Technology of China, Shanghai 201315, P.~R.~China}

\begin{abstract}
Covert communication offers a method to transmit messages in such a way that it is not possible to detect that the communication is happening at all. In this work, we report an experimental demonstration of covert communication that is provably secure against unbounded quantum adversaries. The covert communication is carried out over 10 km of optical fiber, addressing the challenges associated with transmission over metropolitan distances. We deploy the protocol in a dense wavelength-division multiplexing infrastructure, where our system has to coexist with a co-propagating C-band classical channel. The noise from the classical channel allows us to perform covert communication in a neighbouring channel. We perform an optimization of all protocol parameters and report the transmission of three different messages with varying levels of security. Our results showcase the feasibility of secure covert communication in a practical setting, with several possible future improvements from both theory and experiment.
\end{abstract}

\maketitle

Security in communication can take different forms. Encryption provides security by hiding the content of messages from eavesdroppers. Authentication ensures that messages originate from the desired sender, and anonymity allows a person to transmit messages without revealing their identity. In covert communication, security is provided by hiding the fact that the communication is happening at all, which is relevant in any scenario where revealing that communication is occurring can be incriminating to the parties involved. Covertness can serve as a powerful tool for securing communications by adding additional protection to transmitted messages.

There has been significant progress in determining the fundamental limits of covert communication over noisy channels. In particular, a square-root law has been established stating that $O(\sqrt{N})$ covert bits can be reliably sent over $N$ channel uses, while also providing explicit protocols that achieve this rate
\cite{bash2013limits,bash2013quantum,che2013reliable,bash2015hiding,wang2016fundamental,sheikholeslami2016covert,
arumugam2016keyless,bloch2016covert,bash2016covert}. The square-root law has also been tested in a table-top experiment \cite{bash2015quantum}, under strong assumptions about Eve's power. In that experiment, the authors assumed that Eve's detectors have larger dark counts than Bob's and she could only access photons that did not reach Bob instead of intercepting all the signals. Covert communication can be proven to be secure even in the presence of unbounded quantum adversaries \cite{wang2016optimal,bash2015quantum} and recently, covert communication has been extended to the quantum case by showing that quantum communication protocols can be carried out covertly, either through noisy channels as in the classical case \cite{arrazola2016covert} or by exploiting relativistic quantum effects \cite{bradler2016absolutely,bradler2017covert}. Additionally, it was shown in Ref. \cite{arrazola2017secret} that messages sent covertly are also secret, a feature that can be used to perform secret key expansion from covert communication. Despite these advances, there has not yet been an experimental demonstration of covert communication that is secure against an unbounded quantum adversary and that is conducted in the regime of metropolitan network distances.

Meanwhile, current optical data networking systems increase capacity by transmitting multiple data streams simultaneously in different colors of light in a dense wavelength-division multiplexing (DWDM) infrastructure. Any communication system needs to fit into this infrastructure to ensure low cost of deployment. However, there exists crosstalk noise between different DWDM communication channels, which is mainly due to the Raman scattering noise generated in the fiber. In quantum communication, this crosstalk should be avoided, but in covert communication, it can serve as a source of noise for hiding signals. In that sense, covert communication fits naturally in modern optical fiber communication.

In this work, we report the first experimental demonstration of secure covert communication over metropolitan distances in a DWDM infrastructure, where no assumptions are made about the adversary's power. The noise in the channel -- which permits covert communication to take place -- originates from crosstalk from classical communication in a neighbouring DWDM channel. Thus, classical communication between two parties opens the possibility of performing covert communication between the sender and another party. We perform numerical optimization to obtain the noise levels and signal intensities that both maximize security and minimize the total running time of the protocol while ensuring the reliability of the communication. We covertly transmit three distinct messages of varying lengths and security parameters, showcasing the interplay between covertness, message length, and running time. \\

\textit{Covert communication protocol.--} Alice wants to transmit a message to Bob in such a way that Eve cannot detect that they are communicating. To quantify Eve's ability to detect the communication, we assume that Alice is equally likely to communicate or not and Eve's goal is to correctly distinguish between these two scenarios. Eve's detection error probability $P_e$ is given by $P_e=\frac{1}{2}(P_{FA}+P_{MD})$, where $P_{FA}$ is the probability of a false alarm and $P_{MD}$ is the probability of a missed detection. Alice and Bob's goal is to prevent Eve from performing better than a random guess, i.e. they want $P_e\geq \frac{1}{2}-\epsilon$ for sufficiently small $\epsilon>0$. We refer to $\epsilon$ as the \emph{detection bias}.

Covert communication is possible in the presence of noise in the channel linking Alice to Bob. We consider the case where a single bit of information is encoded in subsequent time-bin modes, with the early time-bin encoding a ``0" and the later one encoding a ``1". We further assume that Alice and Bob have access to $N$ such time-bin pairs, each of which may be used to send a signal. Conceptually, the protocol for covert communication is simple: for each of the $N$ time-bin pairs, Alice sends a qubit signal with probability $q\ll 1$, and with probability $1-q$, she does nothing. By doing so, Alice is effectively hiding her signals in the noise by spreading them randomly in time. Alice and Bob require shared randomness to specify the time-bins where signals are sent, which must be kept secret from Eve. The average number of signals sent is $d=qN$, which fixes $q$ for a given message length.

In the absence of communication, the state of each mode is a noisy state $\rho$. When Alice communicates with Bob, the state of each signal mode is given by the mixture $\sigma=(1-q)\rho+q\rho_s$, where $\rho_s$ is the state when Alice sends a signal in that mode. As shown in Refs. \cite{bash2015quantum,arrazola2016covert}, the detection bias can be bounded by the expression
\beq\label{EQ: det bias}
\epsilon\leq \sqrt{\frac{N}{8}D(\rho||\sigma)},
\eeq
where $D(\rho||\sigma)$ is the quantum relative entropy. Thus, in order to calculate an upper bound on the detection bias for a given protocol, we only need to specify the states $\rho$ and $\sigma$ and compute their quantum relative entropy.

It is important to keep in mind that Alice and Bob are not hiding the fact that they are capable of communicating covertly, nor that there is a channel linking them. Instead, they wish to prevent Eve from determining if they are ever talking to each other. This enables them to add increased security to their communications, for example if they fear that their encryption methods may be compromised by intrusive governments with access to trapdoors, or if they want to conceal preparations for an announcement such as a merger of their companies. In the remainder of this paper, we discuss the experimental setup for covert communication and provide a detailed description of the protocol used. We then report results on the covert transmission of different messages in various regimes.

\begin{figure*}
\centering
\resizebox{13cm}{!}{\includegraphics{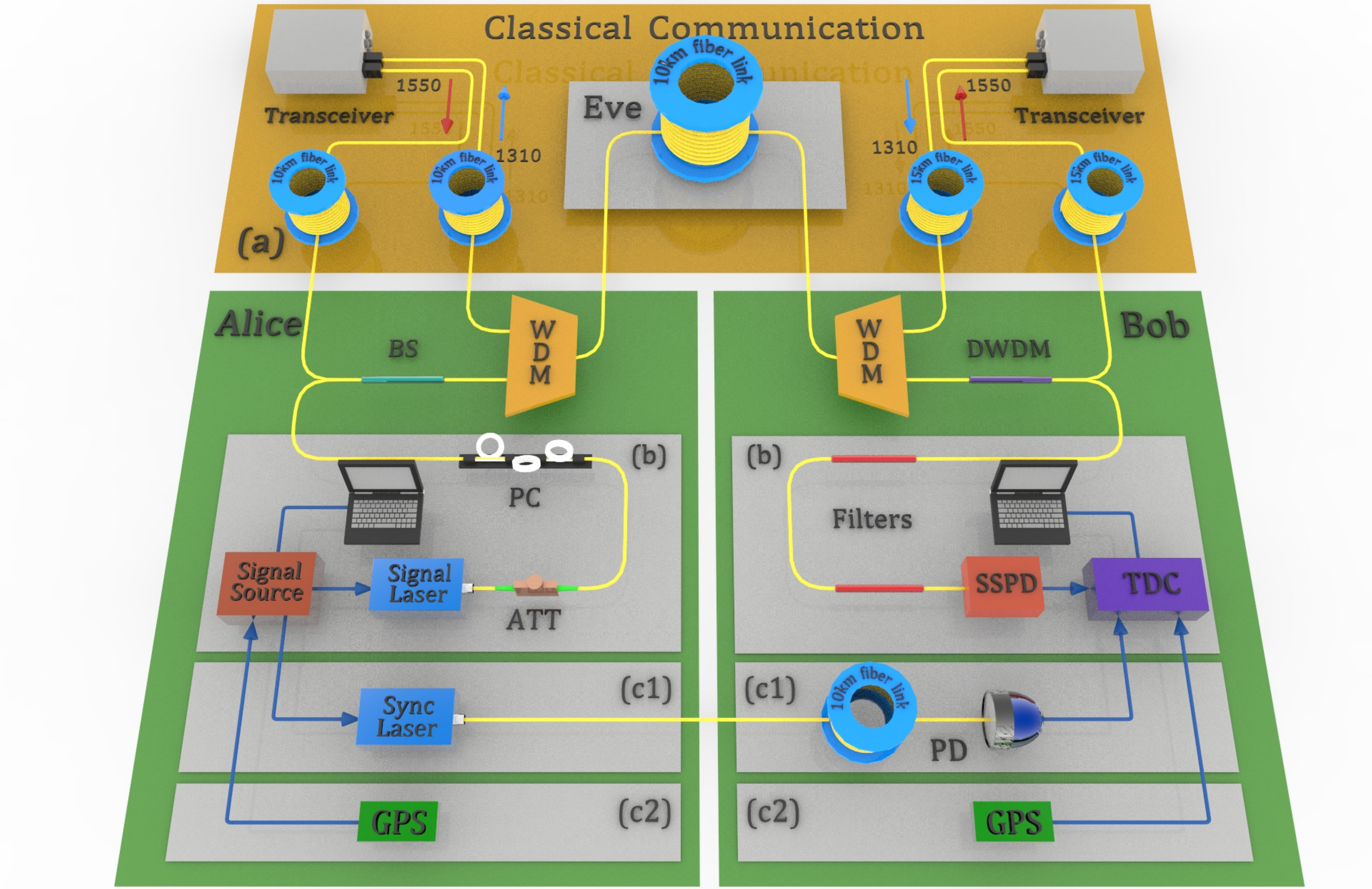}}
\caption{(Color online) Experimental setup for covert communication. (a) Alice sends/receives classical signal in a wavelength $\lambda_1$=1550 nm/$\lambda_2$=1310 nm to and from Bob. The signals transmit in a 10 km fiber spool before combining with signals in other wavelengths, and transmit in a 15 km fiber spool after splitting from other wavelengths. In between Alice and Bob, a 10 km fiber spool is used to transmit signals in all wavelengths. In the experiment, we use commercial fiber transmitters for this classical communication. (b) Alice inserts covert signals of wavelength $\lambda_3$=1560 nm to the same fiber using a beam splitter (BS) and wavelength-division multiplexing (WDM). The positions of the covert signals are calculated with pre-shared quantum random numbers in a personal computer (PC). This is then converted into electrical signals with a signal source, which includes a National Instruments (NI) field-programmable gate array (FPGA) card and subsequently an arbitrary wave generator (AWG). The electric signals trigger a laser diode that sends covert signals, whose amplitude is set with an attenuator (ATT) before inserting them into the fiber spool. Bob uses a DWDM to demultiplex the covert signal from the classical light. Two 1560 nm filters and a 1310/1560 nm WDM are used to eliminate any residual classical light. The covert signal is detected with a superconducting nanowire single-photon detector (SNSPD) and recorded with a time-digital converter (TDC). All data is collected in a PC for analysis. (c1) Sync pulse synchronization. A sync pulse of 100 kHz repetition rate is generated by the signal source, which is then sent to Bob at wavelength $\lambda_4$ = 1532 nm. Bob  records the sync pulse in the TDC for analysis. (c2) GPS synchronization. A global positioning system (GPS) 1 pps signal is used to trigger both Alice's and Bob's system as a starting signal. No sync pulses are needed in this scheme since the timing depends on the internal clock of the electronics.
}\label{Fig:setup}
\end{figure*}

\textit{Experiment.--} Any noise that Eve cannot control can be used to hide the covert signals for covert communication. In fiber communication, Raman scattering always exists when there is classical light travelling due to the fiber's nonlinearity. A single mode fiber supports more than 70 different wavelength channels with DWDM technology. Thus, the free spectral channels with Raman noise can be used for covert communication. Suppose Alice is sending messages at wavelength $\lambda_1$ = 1550 nm in a DWDM network, and receiving messages at wavelength $\lambda_2$ = 1310 nm, as shown in Fig.~\ref{Fig:setup}(a). Through the 10 km fiber before Alice combines the different wavelengths, the Raman noise is generated and covers all the DWDM channels in the same fiber. As a result, noise exists in all other wavelength channels, in our case $\lambda_3$ = 1560 nm, which can be used as the covert communication channel. The experimental setup for covert communication is shown in Fig~\ref{Fig:setup}(b). The covert signal is combined with the classical communication as well as the noise using a beam splitter (BS) and wavelength-division multiplexing (WDM). The noise at Alice's output is measured as $\bar{n}_A$. After transmitting in 10 km fibers, the noise at Bob's input is measured as $\bar{n}_B$. As the Raman noise is generated when the classical light is transmitted, the noise at Bob may be higher than that at Alice. These parameters are used to calculate and optimize the detection bias for covert communication. The covert signal is split from the classical light for detecting. The basic encoding and decoding is processed in a personal computer.

A message with a few characters is decided before the experiment. Each character is encoded with a 5-bit string, allowing 32 possible characters: the 26 letters of the alphabet and a few special characters. Because there are significant losses and there is noise in the channel, it is necessary to perform error-correction. We adopt the strategy of using a repetition code, where the codewords consist of $k$ repetitions of each single bit. For a pure-loss erasure channel with transmissivity $\eta$, the channel capacity is precisely $\eta$, and the repetition code performs well compared to this while allowing significantly simpler encoding and decoding than more sophisticated codes. In this case, the desired decoding error probability, the overall transmissivity of the channel, and the number of bits in the message determine the average number of signals sent and therefore the detection bias of the protocol. To minimize the total running time of the protocol and to maximize security, we fix a target detection bias and perform a numerical optimization to deduce the optimal signal intensities, as specified in detail in the Supplemental Material.

Based on pre-shared random numbers, Alice calculates the positions to send the covert bits. Encoded pulses are generated in signal source, with a National Instruments (NI) field-programmable gate array (FPGA) card and subsequently an arbitrary wave generator (AWG) for coarse and fine timing control. A laser diode finally sends the covert signals to Bob via the DWDM network. The laser diode works at wavelength $\lambda_3$ = 1560 nm with 1 ns pulse width, allowing a maximum repetition rate of 500 MHz in the experiment.

The covert message travels through 10 km of fiber between Alice and Bob, with a transmission loss of around 2.5 dB. A WDM and a DWDM demultiplex the covert signals from the classical signals which are sent to and from the fiber transmitter. One additional WDM and two 1560 nm filters are used to remove any residual light not at the covert wavelength $\lambda_3$ = 1560 nm. The total losses including the WDM and filters are measured to be 3.3 dB. A superconducting nanowire single photon detector (SNSPD) with detection efficiency of 70\% is used to detect the covert signal. The detection results are recorded in a time digital convertor (TDC). Bob records the detections at the covert positions, then recovers the message using a majority vote to decode the transmitted bits.

To synchronize Alice and Bob, a laser diode with repetition rate 100 kHz sends sync signals to Bob, as shown in Fig.~\ref{Fig:setup} (c1). Although this method is useful in an experiment, the sync pulses may leak the information that Alice and Bob are communicating. In order to prevent Eve from using this to detect communication, it is necessary that this pulse is also sent even when no communication takes place. Note that this sync signal is only used for easiness in this experiment, but could be dispensed by using highly accurate clocks. Indeed, we show that it is possible to communicate covertly without sending sync signals. As shown in Fig.~\ref{Fig:setup} (c2), GPS receivers are used as a local clock to trigger Alice's signal source and Bob's recording device. The timing jitter of a GPS signal is measured to be around 30 ns (rms). The repetition rate of the covert signal is then set to 500 kHz to minimize the error between adjacent signals. Note that the repetition frequency using GPS is low because of the large timing jitter. However, by using a GPS assisted atomic clock, it is not hard to achieve a timing jitter of $~10^{-10}$ s, thus enabling a covert signal rate of several gigahertz.

\begin{center}
\begin{table}
\begin{tabular}{ |c | c | c | c| }
\hline
 Message 		& CQTUSTC 	& PRTYSAT@NINE	& QPQI  \\
 \hline
 Bits			& 35			& 60				& 20	  \\
 Detection bias	& 0.014 		& 0.055			& 0.067 \\
 Repetition rate & 500 MHz  & 500 MHz 		& 500 kHz\\
 Sync pulse  	& Yes		& Yes 			& No\\
 Time-bins 		& $1.56\times 10^{12}$	& $2.17\times 10^{11}$	& $3.71\times 10^{9}$ \\
 Covert signals & 68,651					& 96,919		& 8,416 \\
 $\mu$ 			& $3.52\times 10^{-2}$ 	& $3.79\times 10^{-2}$ & 0.266 \\
 $\bar{n}_A$	& $2.30\times 10^{-3}$ 	& $2.50\times 10^{-3}$ & 0.60 \\
 $\bar{n}_B$	& $3.18\times 10^{-3}$ 	& $2.74\times 10^{-3}$ & 0.68 \\
 Running time (s) & 3,120	& 434		& 7,420\\
\hline
\end{tabular}
\caption{Summary of experimental parameters for different covert messages. Here $\mu$ is the mean photon number of the signals, $\bar{n}_A$ and $\bar{n}_B$ are the mean photon number of the noise at Alice's output and Bob's input. }\label{Tab:Params}
\end{table}
\end{center}

\textit{Results.---}
We send three different messages covertly, with different lengths and security. The messages are (i) ``CQTUSTC" -- the acronyms of our academic institutions, (ii) ``PRTYSAT@NINE" -- to model Alice and Bob covertly organizing a surprise party for Eve, and (iii) ``QPQI" -- quantum physics and quantum information. The protocol parameters have been calculated through a numerical optimization routine and are summarized in Table \ref{Tab:Params}. In the optimization, the complete system transmittance is 18\% based on our measurements, and the decoding error probability for each message is set to 1\%. Given the model of the signals and the noise states, a detection bias can be calculated for each covert transmission using Eq. \eqref{EQ: det bias} as shown in the Supplemental Material.

Take the message ``CQTUSTC'' as an example: In our encoding scheme, there are 35 bits to transmit corresponding to seven characters. The security level of $\epsilon=0.014$ and the covert signal repetition rate of 500 MHz are set before the optimization. The noise per time-bin at Alice, given as a mean photon number, is measured to be $\bar{n}_A = 2.30\times 10^{-3}$, and that at Bob is measured to be $\bar{n}_B = 3.18\times 10^{-3}$. The optimization is performed to send the signals spread out in the smallest number of time-bins to fulfill the security requirement. As a result, the mean photon number of the covert pulse is optimized to be $\mu=3.52\times 10^{-2}$, while Alice needs to send 68,651 covert pulses in $1.56\times 10^{12}$ time bins, with each bit repeated 1,961 times. Analogous calculations are done for the other messages.

The detected bits are decoded and error corrected to generate the received message. The statistical results are summarized in Table \ref{Tab:Result}. For instance, in receiving the message ``CQTUSTC'', the detection probabilities are $8.42\times 10^{-3}$ per pulse for a signal mode and $1.04\times 10^{-3}$ per bin for the noise. As each bit in the message sent is repeated 1,961 times, on average 16.5 repeated bits are detected for one original bit. With an average error rate of 13.67\%, all the bits in the original message are decoded correctly with a majority vote decoding, as expected from the protocol. Detailed results of the transmission can be found in Figure \ref{Fig:Data1}. Analogous considerations are done for the other messages, as discussed in the Supplemental Material.

\begin{figure}[htb]
\includegraphics[width=9cm]{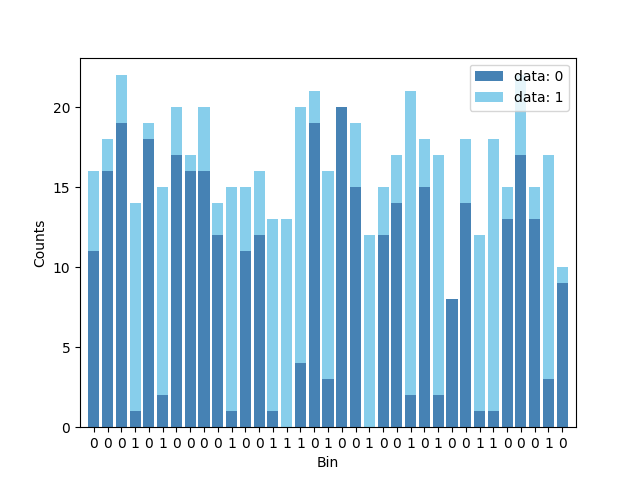}
\caption{(Color online) Results for the covert transmission of the message ``CQTUSTC". The message consists of 35 bits which are all decoded correctly. Each bar shows the total number of ``0" signals (dark blue) and ``1" signals (light blue). On average, $\sim$17 signals were received per bit, with an error rate of $\sim$13.7$\%$.}\label{Fig:Data1}
\end{figure}

\begin{figure}[htb]
\centering
\resizebox{7cm}{!}{\includegraphics{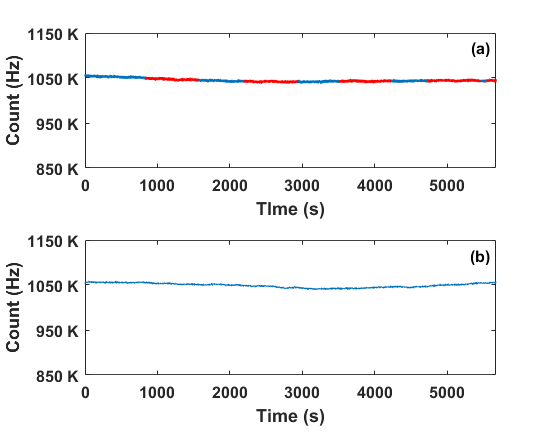}}
\caption{(Color online) Eve's monitoring of the signal rates when attempting to detect the communication. (a) Detection rate when transmitting the message ``CQTUSTC". (b) Detection rate when no communication takes place. Note the blue points indicate Alice is preparing the message, the red points indicate Alice is sending message.
}
\label{Fig:rate}
\end{figure}

Finally, to illustrate the difficulty that Eve faces in detecting the communication, we consider an attack where she monitors the signal rate in the covert wavelength. We record the detection rate per second with Bob's detector, when sending covert signals and when not doing so. The statistical result is shown in Fig.~\ref{Fig:rate}. There is no noticeable change in the intensity and distribution whether Alice sends covert signal, thus illustrating the covertness of the communication.

\textit{Discussion.---} Quantum cryptography has been frequently associated exclusively with quantum key distribution. However, the landscape of cryptography is much larger and covert communication is a promising new research direction in the field: it is a natural problem that arises in many relevant scenarios while providing a form of security beyond encryption. In this work, we have reported the first experimental demonstration of covert communication that is provably secure against an unbounded quantum adversary and that is carried out in a realistic telecommunication scenario, notably in the presence of high losses associated with long distance transmissions.

Several advances in both the theoretical and experimental sides are possible, notably in terms of improving the performance of the protocols, tightening the security bounds, and addressing the range of applicability of covert communication techniques. For example, better error-correction codes could be used to deal with high losses , while developing the necessary software to deploy them. Similarly, Q-ary encodings -- where multiple bits are encoded by sending a signal in one out of Q possible modes -- can be considered. Running times could be reduced by multiplexing signals over many covert channels while also using other degrees of freedom such as polarization and orbital angular momentum to encode information. Finally, it is an intriguing prospect to extend these results to covert quantum communication protocols.
\section*{Supplemental Material}
We provide details of the security calculations, the optimization of parameters and all the steps of the covert communication protocol. We also include the complete data obtained for the message transmission.

\textit{Calculation of signal and noise states.---} The security of the protocol depends on Eve's states $\rho$ when there is no communication and $\sigma$ when there is covert communication. We model the background noise originating from Raman scattering as a thermal state with mean photon number $\bar{n}_A$. In this case, the state $\rho$ is simply given in the Fock basis as
\begin{equation}
\rho = \sum_{n = 0}^{\infty} Q_{\bar{n}_A}(n) \ketbra{n}{n}
\end{equation}
where $Q_{\bar{n}_A}(n)$ is the probability mass function of thermal distribution with mean photon number $\bar{n}_A$ given by
\begin{equation}
Q_{\bar{n}_A}(n) = \frac{\bar{n}_A^n}{(1 + \bar{n}_A)^{n+1}}.
\end{equation}

On the other hand, the state when there is a communication between Alice and Bob $\sigma$ depends on the signal state $\rho_S$. The signals that Alice sends are phase-randomized coherent states with mean photon number $\mu$, which follow a Poisson distribution of photon numbers with probability mass function $P_\mu(n)$. Taking into account the constant background thermal noise, the state seen by Eve when a signal is state in that mode is given by
\begin{equation}
\rho_S = \sum_{n = 0}^{\infty} p(n) \ketbra{n}{n}
\end{equation}
where
\begin{align}
p(n) &= \sum_{r = 0}^{n} P_\mu(r) \times Q_{\bar{n}_A}(n-r)\\
&= \sum_{r = 0}^{n} \frac{e^{-\mu} \mu^r}{r!}  \frac{\bar{n}_A^{n - r}}{(1 + \bar{n}_A)^{n - r + 1}}
\end{align}
The state $\sigma$ is then given by $\sigma = q \rho_S + (1 - q) \rho$ where $q$ is the probability of sending a signal in that mode.\\

The detection bias is then bounded as
\beq\label{EQ: det bias}
\epsilon\leq \sqrt{\frac{N}{8}D(\rho||\sigma)},
\eeq
where $2N$ is the total number of time-bins.

\textit{Decoding error probability. ---}  To calculate the decoding error probability, we first calculate the probability $p_C$ that Bob's detector records a click in the correct mode where a signal is sent. Given a total transmission efficiency $\tau$ in the channel, which includes the limited detector efficiency, the probability of a correct click is given by
\begin{equation}
p_C = 1 -  \frac{\exp(-\tau \mu) }{(1 + \tau\bar{n}_B)}.
\end{equation}
Similarly, the probability of obtaining a click on the wrong mode where there is only noise is given by
\begin{equation}
p_W = 1 - \frac{1}{(1 + \tau \bar{n}_B)}.
\end{equation}

From this, we can calculate the probability $p_g$ that, given there was a click, it is a correct one. Ignoring cases where we observe clicks in both modes, this is given by
\begin{equation}
p_g = \frac{p_C}{p_C + p_W}.
\end{equation}
In the protocol, we use the repetition code to protect against errors and losses, using the majority vote for decoding. Suppose that Alice repeats each bit $k$ times and Bob observes $i$ clicks; then an error occurs when at least $i/2$ of the clicks are in the wrong mode. The probability $\delta$ that Bob incorrectly decodes an individual bit is given by
\begin{align}
\delta = &\sum_{i = 0}^{k} \text{Pr[$i$ clicks]} \times \text{Pr[majority wrong]}\nonumber\\
= &\sum_{i = 0}^{k} \binom{k}{i} (p_C + p_W)^i (1 - p_C - p_W)^{k - i}\times\nonumber\\
 &\sum_{j = 0}^{\lfloor i/2 \rfloor} \binom{i}{j} (p_g)^j (1 - p_g)^{i - j}. \label{eq: error}
\end{align}
To decode Alice's message reliably, Bob has to decode all the bits in the message correctly. For each bit, he can do this with probability $(1 - \delta)$. Thus, the decoding error probability for the entire message $\mathcal{E}$ is given by
\begin{equation}\label{eq: dec. error}
\mathcal{E} = 1 - (1 - \delta)^b
\end{equation}
where $b$ is the number of bits in the message.\\

\textit{Detailed covert communication protocol.---} Suppose Alice and Bob wish to covertly exchange a message of $b$ bits with a detection bias of $\epsilon$ and a decoding error probability $\mathcal{E}$. Then they perform the following protocol:
\begin{enumerate}
\item From Eqs. \eqref{eq: error} and \eqref{eq: dec. error}, calculate the smallest number of signals $d$ that is required to achieve the
    desired decoding error probability. The number of repetitions in the repetition code is $k=d/b$.
\item Fix the probability $q$ of sending a signal in each slot to be $q=d/N$, where $N$ is the total number of time-bin pairs.
\item Fix the mean photon number of the signals to be $\mu$.
\item Using Eq. \eqref{EQ: det bias}, calculate the smallest $N$ such that the desired detection bias $\epsilon$ is reached.
\item Repeat these steps for different values of $\mu$ to find the optimal value that minimizes the total number of time-bins.
\item To transmit the message covertly, use a quantum random number generator to choose in which time-bin pairs to send a signal, each with
    probability $q$. Let $d'$ be the actual number of signals sent.
\item Use $k'=d'/b$ as the number of repetitions of each bit and decode using a majority vote.
\end{enumerate}

\textit{Experimental results. ---} Below, we reproduce the data obtained in the transmission of each message of the covert communication protocol. The results are in Table~\ref{Tab:Result} reporting the signal probability per pulse, noise probability per bin, and error rates. We also illustrate the results in terms of bar graphs, where each bar shows the total number of ``0" and ``1" signals obtained in the transmission of each encoded bit. The message's bit value is decoded by taking the majority vote of the outcomes, resulting in a correct decoding for all messages.

For the message ``QPQI" which operates at a lower repetition rate, the smallest running time is obtained by setting larger noise levels, thus reducing the total number of time-bins while still not affecting the repetition rate. In this case, large signal detection probability compared to noise probability would lead to insecurity, so we must endure higher error rates which are compensated by a larger number of repetitions.

As a comparison, we send the same message using a continuous wave (CW) laser instead of the fiber transmitter. the parameter and the result are shown in Table~\ref{Tab:CWParams} and Table~\ref{Tab:CWResult}.

\begin{center}
\begin{table}
\begin{tabular}{ |c | c | c | c| }
\hline
 Message 		& CQTUSTC 			& PRTYSAT@NINE			& QPQI  \\
 \hline
 Signal probability & $8.42\times 10^{-3}$ & $8.62\times 10^{-3}$	& $9.26\times 10^{-2}$ \\
 Noise probability	& $1.04\times 10^{-3}$ & $9.01\times 10^{-4}$	& $2.23\times 10^{-2}$ \\
 Error rate       & 13.67\% 			 & 14.85\% 				& 23.62\% \\
\hline
\end{tabular}
\caption{Summary of experimental results for different covert messages. Signal probability denote the probability of detecting a click when a signal is sent in a particular mode, while the noise probability denote the probability for bins where there is only noise. The resulting error rates are sufficiently low to recover the correct bits with a majority vote decoding.}\label{Tab:Result}
\end{table}
\end{center}

\begin{figure}[hbt]
\includegraphics[width=9cm]{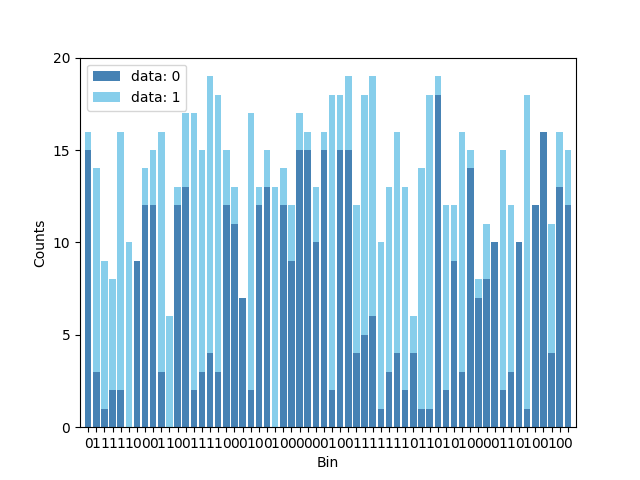}
\caption{(Color online) Results for the covert transmission of the message ``PRTYSAT@NINE". The message consists of 60 bits which are all decoded correctly. Each bar shows the total number of ``0" signals (dark blue) and ``1" signals (light blue). On average, $\sim$14 signals were received per bit, with an error rate of $\sim$14.9$\%$.}\label{Fig:Data2}
\end{figure}

\begin{figure}[hbt]
\includegraphics[width=9cm]{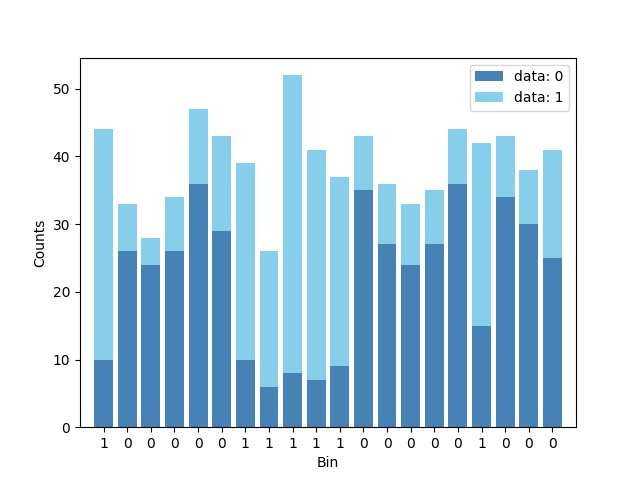}
\caption{(Color online) Results for the covert transmission of the message ``QPQI". The message consists of 20 bits which are all decoded correctly. Each bar shows the total number of ``0" signals (dark blue) and ``1" signals (light blue). On average, $\sim$39 signals were received per bit, with an error rate of $\sim$23.6$\%$.}\label{Fig:Data3}
\end{figure}

\begin{center}
\begin{table}
\begin{tabular}{ |c | c | c | c| }
\hline
 Message 		& CQTUSTC 	& PRTYSAT@NINE	& QPQI  \\
 \hline
 Bits			& 35			& 60				& 20	  \\
 Detection bias	& 0.015 		& 0.069			& 0.070 \\
 Repetition rate & 500 MHz  & 500 MHz 		& 500 kHz\\
 Sync pulse  	& Yes		& Yes 			& No\\
 Time-bins 		& $1.56\times 10^{12}$	& $2.17\times 10^{11}$	& $3.71\times 10^{9}$ \\
 Covert signals & 68,651					& 96,919		& 8,416 \\
 $\mu$ 			& $3.57\times 10^{-2}$ 	& $4.17\times 10^{-2}$ & 0.278 \\
 $\bar{n}_A$	& $2.14\times 10^{-3}$ 	& $2.18\times 10^{-3}$ & 0.62 \\
 $\bar{n}_B$	& $3.48\times 10^{-3}$ 	& $2.93\times 10^{-3}$ & 0.66 \\
 Running time (s) & 3,120	& 434		& 7,420\\
\hline
\end{tabular}
\caption{Summary of experimental parameters for different covert messages using CW laser as noise source instead of the fiber transmitter. Here $\mu$ is the mean photon number of the signals, $\bar{n}_A$ and $\bar{n}_B$ are the mean photon number of the noise at Alice's output and Bob's input. }\label{Tab:CWParams}
\end{table}
\end{center}

\begin{center}
\begin{table}
\begin{tabular}{ |c | c | c | c| }
\hline
 Message 		& CQTUSTC 			& PRTYSAT@NINE			& QPQI  \\
 \hline
 Signal probability & $7.91\times 10^{-3}$ & $9.41\times 10^{-3}$	& $9.31\times 10^{-2}$ \\
 Noise probability    & $8.15\times 10^{-4}$ & $9.54\times 10^{-4}$	& $2.15\times 10^{-2}$ \\
 Error rate       & 13.08\% 			 & 10.40\% 				& 23.34\% \\
\hline
\end{tabular}
\caption{Summary of experimental results for different covert messages using CW laser as noise source instead of the fiber transmitter. Signal probability denote the probability of detecting a click when a signal is sent in a particular mode, while the noise probability denote the probability for bins where there is only noise. The resulting error rates are sufficiently low to recover the correct bits with a majority vote decoding.}\label{Tab:CWResult}
\end{table}
\end{center}

\bibliography{Bibliography}
\bibliographystyle{apsrev}

\end{document}